\documentclass[conference]{IEEEtran}
\IEEEoverridecommandlockouts
\usepackage{algorithm}
\usepackage{algpseudocode}
\usepackage{amsfonts}
\usepackage{bm}
\usepackage{amsmath}
\usepackage{cite}
\usepackage{amssymb}
\usepackage{enumerate}
\usepackage{color,xcolor}
\usepackage{graphicx}
\usepackage{multirow,tabularx}
\usepackage{subfigure}
\usepackage{hyperref}

\makeatletter

\newcommand{\Rmnum}[1]{\expandafter\@slowromancap\romannumeral #1@}
\makeatother

\ifCLASSINFOpdf

\else

\fi

\PassOptionsToPackage{draft}{hyperref}
\begin{document}

\title{A Low-Complexity Gaussian Message Passing Iterative Detector for Massive MU-MIMO Systems}

%\author{\IEEEauthorblockN{Lei Liu, Chau Yuen, Yong Liang Guan and Ying Li }
%
%\thanks{Lei Liu and Ying Li are with the State Key Lab of Integrated Services Networks, Xidian University, Xian, 710071, China (e-mail: lliu\_0@stu.xidian.edu.cn, yli@mail.xidian.edu.cn).}
%\thanks{Chau Yuen is with the Singapore Unversity of Technology and Design, Singapore (e-mail: yuenchau@sutd.edu.sg).}
%\thanks{Yong Liang Guan is with the School of Electrical and Electronic Engineering, Nanyang Technological University, Singapore (e-mail: yuenchau@sutd.edu.sg).}
%}

\author{\IEEEauthorblockN{Lei Liu\IEEEauthorrefmark{1}, Chau Yuen\IEEEauthorrefmark{2}, Yong Liang Guan \IEEEauthorrefmark{3}, Ying Li\IEEEauthorrefmark{1} and Yuping Su\IEEEauthorrefmark{1}\\
\IEEEauthorrefmark{1}State Key Lab of ISN, Xidian University, Xi'an 710071, China\\
\IEEEauthorrefmark{2}Singapore University of Technology and Design, Singapore\\
\IEEEauthorrefmark{3}Nanyang Technological University, Singapore\\
E-mail:yli@mail.xidian.edu.cn}

\thanks{This work was supported in part by the 973 Program under Grant 2012CB316100 and the National Natural Science Foundation of China under Grant 61301177. The first author was also supported by the China Scholarship Council under Grant 20140690045.}}

\maketitle

%\vspace{-2.5cm}
\begin{abstract}
This paper considers a low-complexity Gaussian Message Passing Iterative Detection (GMPID) method over a pairwise graph for a massive Multiuser Multiple-Input Multiple-Output (MU-MIMO) system, in which a base station with $M$ antennas serves $K$ Gaussian sources simultaneously. Both $K$ and $M$ are large numbers and we consider the cases that $K<M$ in this paper. The GMPID is a message passing algorithm based on a fully connected loopy graph, which is well known that it is not convergent in some cases. In this paper, we first analyse the convergence of GMPID. Two sufficient conditions that the GMPID converges to the Minimum Mean Square Error (MMSE) detection are proposed. However, the GMPID may still not converge when $K/M>(\sqrt{2}-1)^2$. Therefore, a new convergent GMPID with equally low complexity called SA-GMPID is proposed, which converges to the MMSE detection for any $K< M$ with a faster convergence speed. Finally, numerical results are provided to verify the validity and accuracy of the proposed theoretical results.
\end{abstract}
%the sufficient conditions are not satisfied, i.e.,
%\begin{IEEEkeywords}
%Convergence, Gaussian message passing, Gaussian belief propagation, graph-based detection, loopy factor graph, low-complexity MIMO detection, sum-product algorithm.
%\end{IEEEkeywords}

\IEEEpeerreviewmaketitle
\section{Introduction}
Multiuser Multiple-Input and Multiple-Output (MU-MIMO) has become a key technology for wireless communication standards like IEEE 802.11n (Wi-Fi), IEEE 802.11ac (Wi-Fi), WiMAX (4G) and Long Term Evolution (4G). Recent research investigations\cite{Argas2013,Rusek2013} show that MU-MIMO will play a vital role in the future wireless systems. Recently, the massive MU-MIMO, where the Base Station (BS) has a very large number of antennas (e.g., hundreds or even more), has attracted more and more attention \cite{Argas2013, Rusek2013, Marzetta2010, Ngo2012, Dai2013}. For instance, several researchers prove that the massive MU-MIMO can bring huge improvement both in throughput and energy efficiency \cite{Marzetta2010,Ngo2012,Dai2013}.

One of the current challenging problems in massive MU-MIMO is the low-complexity signal detection in the up-link\cite{Rusek2013}. In the case of Gaussian sources, it is well known that Minimum Mean Square Error (MMSE) detection is optimal. However, the complexity is significantly high for its unfavorable matrix inversion \cite{tse2005}. Another kind of MU-MIMO detection is graph based detection called message passing algorithm \cite{ kschischang2001, Loeliger2004}. There are two kinds message passing algorithms, One of which is Gaussian Belief Propagation (GaBP) algorithm based on a graph that consisted by variable nodes \cite{Weiss20011, Weiss20012, malioutov2006, Gao2014, Su2014, moallemi2009} and the other is Gaussian Message Passing Iterative Detection (GMPID) based on a pairwise graph that consists of variable nodes and sum nodes \cite{Guo2008, andrea2005, Roy2001, William2009, wu2014, yoon2014}. Both of them are efficient distributed message passing algorithms for Gaussian graphical models. %GaBP has many applications, such as solving linear equations \cite{moallemi2009, Loeliger2002, Loeliger2006}, signal processing \cite{kschischang2001, Loeliger2002,Loeliger2004,Loeliger2006}, multiuser detection, linear programming, ranking in social networks, support vector machines etc. \cite{Bickson2008, bickson2009}.
Moreover, the GMPID has also been studied for equalization in the inter-symbol interference channel \cite{Guo2008} and the decoding of modern channel codes, such as turbo codes and low density parity check codes \cite{William2009}.

It is proved that if the factor graph is of tree structure, the means and variances of the message passing algorithm converge to the true marginal means and approximate marginal variances respectively \cite{kschischang2001,Loeliger2004}. However, if the graph has cycles, the message passing algorithm may fail to converge. Most previous works of the message passing algorithm focus on the convergence of the GaBP algorithm. Three general sufficient conditions for convergence of GaBP in loopy graphs are known: diagonal-dominance \cite{Weiss20011,Weiss20012}, convex decomposition \cite{moallemi2009} and walk-summability \cite{malioutov2006}. Recently, a necessary and sufficient variances convergence condition of the GaBP is given in \cite{Su2014}. For the GMPID based on the pairwise graph, a sufficient condition of the means convergence is given in \cite{Roy2001}. However, the posterior density matrixes of each sum node are needed to calculate \cite{Roy2001}, which introduces the matrix inversion operation and a much higher computational complexity during the message updating. Montanari \cite{andrea2005} has proved the GMPID algorithm converges to the optimal MMSE solution for any arbitrarily loaded randomly-spread CDMA system, but this only works for CDMA MIMO system with binary channels. In general, the GMPID has lower computational complexity and a better Mean Square Error (MSE) performance than the GaBP algorithm. To the best of our knowledge, the convergence of GMPID based on a pairwise graph is far from completion.

In this paper, we analyse the convergence of the existing GMPID and propose a new convergent detection method for massive MU-MIMO systems with $K$ users and $M$ antennas. Let $\beta  = {K \mathord{\left/
 {\vphantom {K M}} \right.
 \kern-\nulldelimiterspace} M}$ and $\beta<1$. The contributions of this paper are summarized as follows.

 \noindent
\hangafter=1
\setlength{\hangindent}{2em} 1) We prove that the variances of GMPID definitely converge to the MSE of MMSE detection, which also gives an alternative way to estimate the MSE performance for the MMSE detection.

 \noindent
\hangafter=1
\setlength{\hangindent}{2em} 2) Two sufficient conditions are derived to guarantee that the means of GMPID converge to that of MMSE detection for $\{\beta: 0<\beta<(\sqrt{2}-1)^{2}\}$.

 \noindent
\hangafter=1
\setlength{\hangindent}{2em} 3) A new convergent GMPID called SA-GMPID is proposed, which converges and converges faster than GMPID to the MMSE detection for any $\{\beta: 0<\beta<1\}$.

%This paper is organized as follows. In Section II, the MU-MIMO system model and the MMSE detector is introduced. The GMPID is elaborated in Section III, followed by a new fast converge detector SA-GMPID in Section IV. Numerical results are shown in Section V, and we end with conclusions in Section VI.

\section{System Model And MMSE Detector}
In this section, the massive MU-MIMO system model and some preliminaries about the optimal MMSE detector for the massive MU-MIMO systems will be introduced.
\subsection{System Model}
\begin{figure}[ht]
  \centering
  \includegraphics[width=5cm]{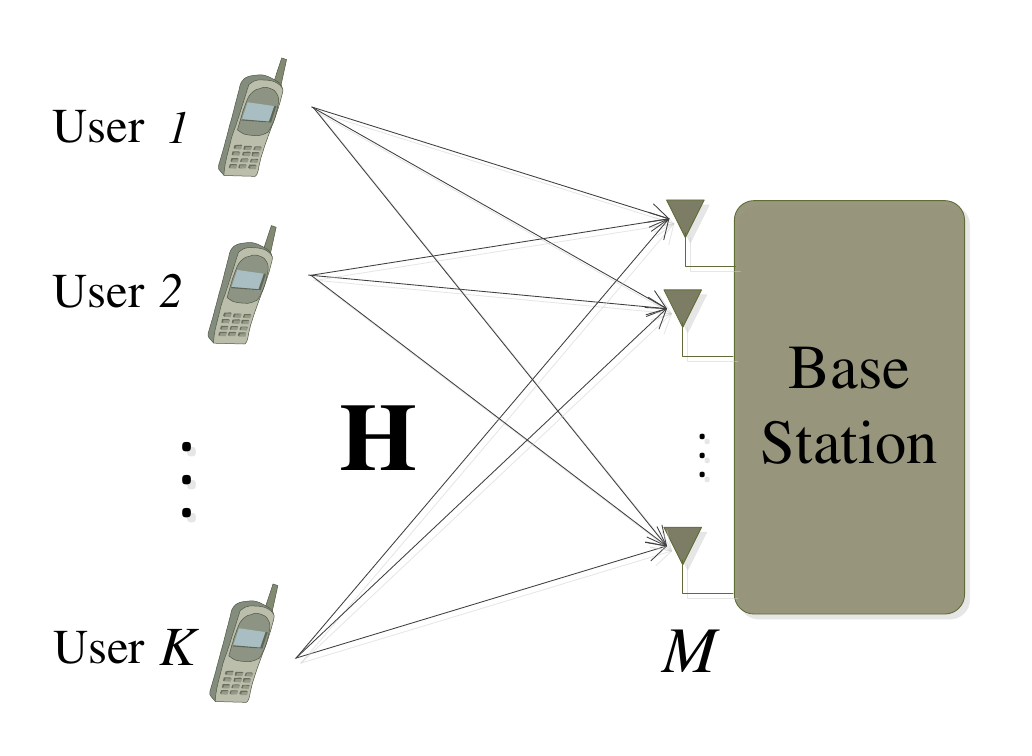}\\
  \caption{MU-MIMO system model: $K$ autonomous single-antenna terminals simultaneously communicate with an array of $M$ antennas of the base station..}\label{f1}
\end{figure}
Fig. \ref{f1} shows the system model. we consider a uplink MU-MIMO system with $K$ users and one BS with $M$ antennas \cite{Marzetta2010,Ngo2012,Rusek2013}. For massive MIMO, the $K$ and $M$ are very large (be hundreds or thousands), e.g., $M=600$ and $K=100$. The $M\times1$ received signal vector ${\large{\textbf{\emph{y}}} }$ at the BS is represented by
\begin{equation}\label{e1}
{\large{\textbf{\emph{y}}} }= \mathbf{H}\textbf{\emph{x}} + \textbf{\emph{n}},
\end{equation}
where $\mathbf{H}$ denotes the $M \times K$ channel matrix, $\textbf{\emph{n}}\sim\mathcal{N}^{\small{M}}(0,\sigma_n^2)$ is an $M \times 1$ independent Additive White Gaussian Noise (AWGN) vector and $\large\textbf{\emph{x}}$ is the message vector sent from $K$ users. We assume that the channels only suffer from the small-scale fading without large-scale fading, in which $\mathbf{H}$ denotes the Rayleigh fading channel matrix whose entries are independently and identically distributed (i.i.d.) with zero mean and unit variance, i.e., normal distribution $\mathcal{N}(0,1)$. Each component of $\large\textbf{\emph{x}}$ is Gaussian distributed, i.e., ${x_k}\sim\mathcal{N}(0,\sigma_{x_k}^2)$, $k \in \{ 1,2,\cdots,K\} $. The task of  multi-user detection at the BS is to estimate the transmitted signal vector ${\large{\textbf{\emph{x}}} }$ from the received signal vector ${\large{\textbf{\emph{y}}} }$. Noting that the channel matrix $\mathbf{H}$ can be usually obtained by time-domain and/or frequency-domain training pilots \cite{Dai2013}, we assume that the BS knows the Channel State Information (CSI) $\mathbf{H}$. In this paper, we only consider the real MU-MIMO system because the complex case can be easily extended from the real case \cite{Gao2014}.

%Minimum mean square error (MMSE) linear detection is suboptimal for the multi-user large-scale MIMO system with the discrete (or digital) sources \cite{Rusek2013}. And the maximum likelihood detection is optimal but NP-hard for multi-user MIMO system \cite{verdu1998}. However, for the sources are gaussian distributed, MMSE detection is the optimal under MSE measure because it minimizes the MSE between sources and estimation. As it will be seen the classical MMSE detection for multi-user MIMO system needs inverse matrix calculation and matrix multiplication. Thus, for the large-scale MIMO system, the complexity of classical MMSE detection is very high. This motivates us to use a low complexity method to do the detection.

\subsection{MMSE Detector}
It is well known that MMSE detection is optimal under MSE measure when the sources are gaussian distributed \cite{verdu1998}. Let $\mathbf{V}_{\textbf{\emph{x}}}$ denote the covariance matrix of the sources $\textbf{\emph{x}}$. Then, the MMSE detector \cite{tse2005} is given by
\begin{equation}\label{GMP2}
{{\hat {\textbf{\emph{x}}}}} = \sigma _{{{n}}}^{ - 2} \mathbf{V} _{{{\hat {\textbf{\emph{x}}}}}}\mathbf{H}^T\textbf{\emph{y}}= \sigma _{{{n}}}^{ - 2}(\sigma _{{{n}}}^{- 2}\mathbf{H}^T\mathbf{H}+\mathbf{V} _{{{ {\textbf{\emph{x}}}}}}^{-1})^{-1}\mathbf{H}^T\textbf{\emph{y}},
\end{equation}
where $\mathbf{V} _{{{\hat {\textbf{\emph{x}}}}}} = (\sigma _{{{n}}}^{- 2}\mathbf{H}^T\mathbf{H}+\mathbf{V} _{{{ {\textbf{\emph{x}}}}}}^{-1})^{-1}$, which denotes the deviation of the estimation to the initial sources. Moreover, the $k$th diagonal element $v_{kk}$ of the covariance matrix $\mathbf{V}_{\hat {\textbf{\emph{x}}}}$ denote the deviation of the estimation to the source $x_k$.

The following is given by the random matrix theory \cite{verdu1998}.
%\begin{equation}\label{PA1}
%\frac{1}{K}{\rm{tr}}\left\{ {{{(\mathbf{I}_K + \gamma {\mathbf{H}^T}\mathbf{H})}^{ - 1}}} \right\} \to 1 - \frac{{\mathcal{F}({\gamma M },\beta )}}{{{{4\gamma \beta M} }}},
%\end{equation}
%where
%\begin{equation}\label{PA2}
%\mathcal{F}(x,z) = {\left( {\sqrt {x{{\left( {1 + \sqrt z } \right)}^2} + 1}  - \sqrt {x{{\left( {1 - \sqrt z } \right)}^2} + 1} } \right)^2}.
%\end{equation}
%
%From (\ref{GMP2}), the MSE of the MMSE detection is calculated by
%\begin{equation}\label{PA5}
%MSE=\frac{1}{K}{\rm{tr}}(\mathbf{V}_{\hat {\textbf{\emph{x}}}}) = \frac{1}{K}{\rm{tr}}\{(\sigma _{{{n}}}^{ - 2}\mathbf{H}^T\mathbf{H}+\mathbf{V} _{{{ {\textbf{\emph{x}}}}}}^{-1})^{-1}\}.
%\end{equation}
%Assume $\mathbf{V}_{\textbf{\emph{x}}}=\sigma_x^2\mathbf{I}_K$ and with (\ref{PA1}), (\ref{PA2}) and (\ref{PA5}) we can get the following proposition.

\textbf{\textit{Proposition 1:}} \emph{When $\beta= K/M<1$ is fixed, $K\rightarrow \infty$ and the sources are i.i.d. with $\mathcal{N}(0,\sigma_x^2)$, the MSE performance of the MMSE detection is described by
\begin{equation}\label{PA6}
MSE = \sigma _x^2\left( {1 - \frac{1}{{4s\beta M}}F(sM,\beta )} \right) \to \frac{{\sigma _n^2}}{{M - K}},
\end{equation}
where $s={{\sigma _x^2} \mathord{\left/
 {\vphantom {{\sigma _x^2} {\sigma _n^2}}} \right.
 \kern-\nulldelimiterspace} {\sigma _n^2}}$ is signal-to-noise ratio.}

When $\beta= K/M<1$, $K\rightarrow \infty$, the MSE of the MMSE detection is determined by the variance of the Gaussian noise and $M-K$, but independent of the variances of the sources.

\textbf{\textit{Remark:}} {The complexity of MMSE detector is $\mathcal{O}(K^3+MK^2)$, where $\mathcal{O}(K^3)$ is for the inverse calculation and $\mathcal{O}(MK^2)$ is for the matrix multiplication $\mathbf{H}^T\mathbf{H}$. As this complexity is very high, it movtivates us to use a low complexity detection method.}

\section{Gaussian Message Passing Iterative Detection}
\begin{figure}[ht]
  \centering
  \includegraphics[width=8cm]{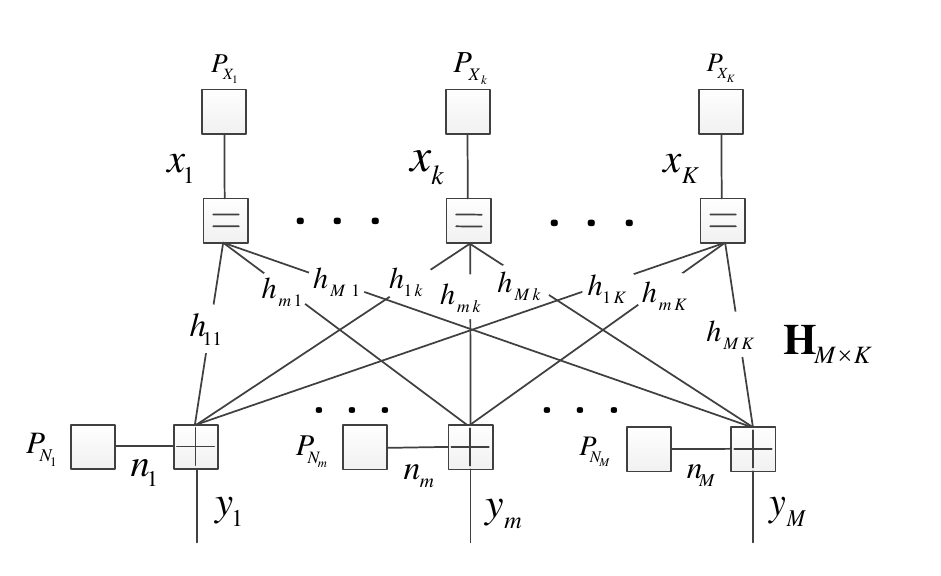}\\
  \caption{Gaussian message passing iterative detection for MU-MIMO systems. The channel parameter from user $k$ to anntenna $m$ is $h_{mk}$, the distribution constranits of each source and noise are denoted by $P_{x_{k}}$ and $P_{N_{m}}$.  }\label{f2}
\end{figure}
In this paper, we consider the GMPID based on a pairwise factor graph for the MU-MIMO system. Fig. \ref{f2} gives the factor graph of MU-MIMO system. The process is very similar to the Belief Propagation (BP) decoding process of LDPC code \cite{William2009}. The differences are: 1) the different passing messages on each edge (the mean and variance of a Gaussian distribution); 2) the different message update functions. %Fig. \ref{f3} presents the message updating diagram. The message updating rules are given as follows.
%\begin{figure}[ht]
%  \centering
%  \includegraphics[width=8.2cm]{update.pdf}\\
%  \caption{Message update at sum nodes and variable nodes. The output message called extrinsic information on each edge is calculated by the messages on the other edges that are connected with the same node. The meassages passing on each edge are the mean and variance of a Guassian distribution.}\label{f3}
%\end{figure}

\subsection{Message Update at Sum Nodes}
Each sum node can be seen as a multiple-access process and the message update at the sum nodes is given by
\begin{equation}\label{e2}
\left\{ \begin{array}{l}
e_{m \to k}^s(t) = {y_m} - \sum\limits_{i \ne k} {h_{mi}e_{i \to m}^v(t-1)\;,} \\
v_{m \to k}^s(t)= \sum\limits_{i \ne k} {h_{mi}^2v_{i \to m}^v(t-1) + \sigma _n^2\;},
\end{array} \right.
\end{equation}
where $i,k \in \{ 1,2, \cdots ,K\} ,m \in \{ 1,2, \cdots ,M\}$, $y_m$ is the $m$-th element of the received vector $\textbf{\emph{y}}$, $h_{mi}$ is the element of channel matrix $\mathbf{H}$ and $\sigma^2_n$ is the variance of the Gaussian noise. What's more, $e_{k \to m}^v(t)$ and $v_{k \to m}^v(t)$ denote the mean and variance passed from $k$th variable node to $m$th sum node respectively, $e_{m \to k}^s(t)$ and $v_{m \to k}^s(t)$ denote the mean and variance passed from $m$th sum node to $k$th variable node respectively. The initial value $\mathbf{v}^v(0)$ equals to $+\boldsymbol{\infty}$ and $\mathbf{e}^v(0)$ equals to $\mathbf{0}$, where $\mathbf{v}^v(t)$ and $\mathbf{e}^v(t)$ are vectors containing the elements $v_{k\to m}^v(t)$ and $e_{k\to m}^v(t)$ respectively.

\subsection{Message Update at Variable Nodes}
Each variable node can be seen as a broadcast process and the message update at the variable nodes is denoted by
\begin{equation}\label{e3}
\left\{ {\begin{array}{*{20}{l}}
{v_{k \to m}^v(t) = {{\left( {\sum\limits_{i } {h_{ik}^2v_{i \to k}^{s{\;^{ - 1}}}(t) + \sigma _{{x_k}}^{ - 2}\;} } \right)}^{ - 1}},}\\
{e_{k \to m}^v(t) = v_{k \to m}^v(t)\sum\limits_{i } {{h_{ik}}v_{i \to k}^{s{\;^{ - 1}}}(t)e_{i \to k}^s(t)\;\;} }.
\end{array}} \right.\quad
\end{equation}
where $k \in \{ 1,2, \cdots ,K\}, i,m \in \{ 1,2, \cdots ,M\}$ and $\sigma^2_{x_k}$ denotes the variance of the source $x_k$.

After the given number of iterations between (\ref{e2}) and (\ref{e3}), we output $\hat{x}_{k}$ as the estimation of $x_k$ and its MSE $\sigma^2_{\hat{x}_{k}}$.
\begin{equation}\label{e4}
{\sigma _{{{\hat x}_k}}^2 = v_{k \to m}^v(t),\quad {{\hat x}_k} = e_{k \to m}^v(t),}
\end{equation}
where $k \in \{ 1,2, \cdots ,K\}$.

It is easy to calculate the computational complexity for each iteration. In each iteration, it needs about $8KM$ multiplications. Therefore, the complexity is low ($\mathcal{O}(KMN_{ite})$), where $N_{ite}$ is the number of iterations.

%\subsection{Comparation with the classical iterative methods}
%The complexity for each iteration of the above classical iterative method is $O(K^2)$, and matrix calculation $A=I_K+snrH^HH$ costs $O(MK^2)$ operations before the iteration. So the total complexity is $O(MK^2+K^2N_{ite})$. When $M$ and $K$ are large enough, the complexity is much higher than the GMPID. The MSE performance comparation will be given in the simulation results. We will see that the proposed GMPID converges faster and always has a better MSE performance than the other classical iterative methods.
%
%\begin{table*}[!t]
%\renewcommand{\arraystretch}{1.8}
%\caption{Complexity comparation of the iterative detection methods for MU-MIMO system}
%\label{table2}
%\centering
%\begin{tabular}{|c|c|c|c|c|}
%\hline
% Detection Methods & Jacobi & $ Richarson $ & GaBP & GMPID \\
%\hline
%  Complexity & $\mathcal{O}(MK^2+K^2N_{ite})$ & $\mathcal{O}(MK^2+K^2N_{ite})$ & $\mathcal{O}(MK^2+K^2N_{ite})$ & $\mathcal{O}(N_{H}N_{ite})$ \\
%\hline
%
%\end{tabular}
%\end{table*}

\subsection{Variances Convergence of GMPID (New Results)}
In this subsection, we give the variances convergence of GMPID. The inequations in this paper correspond to the component-wise inequality. From (\ref{e2})(\ref{e3}), we have
\begin{equation}\label{p1}
{v_{k \to m}^v(t) \!=\! {{\left( \!{\sum\limits_{i } {h_{ik}^2{{\left( {\sum\limits_{j \ne k} {h_{ij}^2v_{j \to i}^v(t-1) + \sigma _n^2\;} } \!\!\right)}^{ - 1}} \!\!\!\!\!\!\!+ \sigma _{{x_k}}^{ - 2}\;} } \!\!\right)}^{ - 1}}}\!\!\!\!\!\!.
\end{equation}
As the initial value $\mathbf{v}^v(0)$ is equal to $+\boldsymbol{\infty}$, it is easy to see that $\mathbf{v}^v(t) > 0$ for any $t>0$ during the iteration. So $\mathbf{v}^v(t)$ has a lower bound $\mathbf{0}$. From (\ref{p1}), we can see that $\mathbf{v}^v(t)$ is a monotonically non-decreasing function with respect to $\mathbf{v}^v(t-1)$. Moreover, we can get $\mathbf{v}^v(1)<\mathbf{v}^v(0)=+\boldsymbol{\infty}$ for the first iteration. Therefore, it can be shown that $\mathbf{v}^v(t)\leq \mathbf{v}^v(t-1)$ with $\mathbf{v}^v(1) \leq \mathbf{v}^v(0)$ from the monotonicity of the iteration function. These mean that $\{\mathbf{v}^v(t)\}$ is a monotonic decreasing sequence but lower bounded. Thus, sequence $\{\mathbf{v}^v(t)\}$ converges to a certain value, i.e., $\mathop {\lim }\limits_{t \to  \infty } {\mathbf{v}^v}(t)=\mathbf{v}^*$.

To simplify the calculation, we assume $\mathbf{V}_{\textbf{\emph{x}}}=\sigma_x^2\mathbf{I}_K$, i.e., $\sigma_{x_k}^2=\sigma_x^2$. With the symmetry of all the elements of $\mathbf{v}^*$, we can get $v_{k\to m}^*=\hat{\sigma}^2,k \in \{ 1, \cdots ,K\}$ and $m \in \{ 1, \cdots ,M\}$. Thus, from  (\ref{p1}), the convergence point $\hat{\sigma}^2$ can be solved by
\begin{equation}\label{p2}
{\hat{\sigma}^2 = {{\left( {\sum\limits_{i } {h_{ik}^2{{\left(\hat{\sigma}^2 {\sum\limits_{j \ne k} {h_{ij}^2 + \sigma _n^2\;} } \right)}^{ - 1}} + \sigma _{{x}}^{ - 2}\;} } \right)}^{ - 1}}}.
\end{equation}
As the channel parameters $h_{ik}^2$ and $h_{ij}^2$ are independent with each other, the above expression can be rewritten as
\begin{equation}\label{p3}
\sigma _{{x}}^{ - 2}\sum\limits_{j \ne k} {h_{ij}^2\;} {{\hat \sigma }^4} + (\sigma _n^2\sigma _{{x}}^{ - 2} + \sum\limits_{i} {h_{ik}^2\; - \sum\limits_{j \ne k} {h_{ij}^2\;} } ){{\hat \sigma }^2} - \sigma _n^2 = 0.
\end{equation}
When $M$ is large, taking an expectation for (\ref{p3}) with respect to the channel parameters $h_{ik}^2$ and $h_{ij}^2$, we get
\begin{equation}\label{p4}
K\sigma _{{x}}^{ - 2}{{\hat \sigma }^4} + (\sigma _n^2\sigma _{{x}}^{ - 2} + M - K ){{\hat \sigma }^2} - \sigma _n^2 = 0.
\end{equation}
Then $\hat{\sigma}^2$ is the positive solution of (\ref{p4}), i.e.,
\begin{equation}\label{p42}
\!\!{{\hat \sigma }^2}\!\! =\!\! \frac{{\sqrt {{{\!(\sigma _n^2\sigma _x^{ - 2} \!+ \!M\! -\! K)}^2}\!\!\! +\! 4K \sigma _x^{ - 2}\sigma _n^2} \! - \!(\sigma _n^2\sigma _x^{ - 2} \!+ \!M\! - \!K)}}{{2K \sigma _x^{ - 2}}}.
\end{equation}
With (\ref{p42}), (\ref{e3}) and (\ref{e4}), we can get the following proposition.

\textbf{\textit{Proposition 2:}} \emph{When $\beta= K/M<1$ is fixed, $K\rightarrow \infty$ and the sources are i.i.d. with $\mathcal{N}(0,\sigma_x^2)$, the variances of GMPID converge to
\begin{equation}\label{p6}
\sigma _{\hat x}^2= {\hat \sigma ^2}\approx
\frac{{\sigma _n^2}}{{M - K + {s^{ - 1}}}}\mathop {}\limits_{\mathop {}\limits_{} } .
\end{equation}}

Comparing (\ref{PA6}) and (\ref{p6}), we can see that it gets the same results as performance analysis by the random matrix theory. Thus, the following theorem can be given.

\textbf{\textit{Theorem 1:}} \emph{When $\beta= K/M<1$ is fixed, $K\rightarrow \infty$ and the sources are i.i.d. with $\mathcal{N}(0,\sigma_x^2)$, the variances of GMPID converge to the exact MSE of the MMSE detection.}

It should be pointed out that the above analysis provides an alternative method to estimate the MSE performance of the MMSE detection. The variances convergence analysis here is simpler and even suitable for the irregular channel matrix. %As we know, if the channel matrix $H$ is irregular, the random matrix theory is hard to give the performance analysis of the MU-MIMO MMSE detection. In the next, simulation results also provide accurate estimations of the MU-MIMO MMSE detection without calculating the large matrix inversion. Another point should be mentioned is that similar to the estimation of random matrix theory, although this result is derived for the large $K$ and $M$ cases, our simulation results shows that it also works well for the small $K$ and $M$ cases.

Similarly, sequence $\{\mathbf{v}^s(t)\}$ also converges to a certain value, i.e., $v^s_{m\to k} \to \tilde{\sigma}^2$. From (\ref{e2}), we can get
\begin{equation}\label{p7}
{\tilde \sigma ^2} \approx K{\hat \sigma ^2} + \sigma _n^2.
\end{equation}
Let $\gamma={{{{\hat \sigma }^2}} \mathord{\left/
 {\vphantom {{{{\hat \sigma }^2}} {{{\tilde \sigma }^2}}}} \right.
 \kern-\nulldelimiterspace} {{{\tilde \sigma }^2}}}$, from (\ref{p6}) and (\ref{p7}), we get
 \begin{equation}\label{p8}
 \gamma  = \frac{1}{{{{K}} + {{\sigma _n^2} \mathord{\left/
 {\vphantom {{\sigma _n^2} {{{\hat \sigma }^2}}}} \right.
 \kern-\nulldelimiterspace} {{{\hat \sigma }^2}}}}} \approx
{\left( {M + {s^{ - 1}}} \right)^{ - 1}}.
 \end{equation}

\subsection{Means Convergence of GMPID (New Results)}
In this subsection, the means convergence of GMPID will be given. Unlike the variances, the means are not always convergent. Two sufficient conditions for the means convergence are given as follows.

\textbf{\textit{Theorem 2:}} \emph{When $\beta= K/M<1$ is fixed and $K\rightarrow \infty$, the GMPID converges to the MMSE estimation if any of the following conditions holds.}

\emph{1. The matrix $\mathbf{I}_K + \gamma\left({\mathbf{H}^T}\mathbf{H}-\mathbf{D}_{{\mathbf{H}^T}\mathbf{H}}\right)$ is strictly or irreducibly diagonally dominant,}

\emph{2. $\rho \left( {\gamma ({\mathbf{H}^T}\mathbf{H} - {\mathbf{D}_{{\mathbf{H}^T}\mathbf{H}}})} \right) < 1$.}

\emph{\!\!\!\!\!\!Where  $\gamma={{{{\hat \sigma }^2}} \mathord{\left/
 {\vphantom {{{{\hat \sigma }^2}} {{{\tilde \sigma }^2}}}} \right.
 \kern-\nulldelimiterspace} {{{\tilde \sigma }^2}}}$ and $ \rho \left( \mathbf{A} \right)$ is the spectral radius of $\mathbf{A}$.}

\begin{IEEEproof}
The proof is omitted due to the page limit.
\end{IEEEproof}

As $\gamma \to \frac{1}{M+s^{-1}}$ and $K \infty$ with $\beta<1$, from random matrix theory, we have
\begin{equation}\label{radius1}
\rho(\gamma(\mathbf{H}^T\mathbf{H}-\mathbf{D}_{\mathbf{H}^T\mathbf{H}}))\to \beta+2\sqrt{\beta},
\end{equation}
for a finite $s$. Then, from the second condition of Theorem 2, we have the following corollary.

\textbf{\emph{Corollary 1:}}  \emph{When $\beta= K/M<1$ is fixed and $K\rightarrow \infty$, the GMPID converges to the MMSE estimation if $\beta<(\sqrt2-1)^2$.}

%Let $\bigtriangleup E(t)=E^*-E^v(t)$ be the means deviation vector. From (\ref{p11}), we can get
%\begin{equation}\label{conv_speed}
%\Delta E(t) = \gamma \left( {{H^H}H - {D_{{H^H}H}}} \right)\Delta E(t - 1).
%\end{equation}
%Therefore, the means converge with an exponential rate of the spectral radius $\rho \left( {\gamma ({H^H}H - {D_{{H^H}H}})} \right)$, i.e., the smaller spectral radius is, the faster convergence it will be.

\section{A New Fast Converge Detector SA-GMPID}
From Corollary 1, we know that when $(\sqrt2-1)^2\leq\beta<1$ the GMPID may not converge. So, in this subsection, we will give a \emph{scaled-and-added} GMPID called SA-GMPID to fix the convergence of GMPID. In the following, let ${\mathbf{H}'}=\sqrt{w}\mathbf{H}$ and $\textbf{\emph{y}}'=\sqrt{w}\textbf{\emph{y}}$ , where $h'_{mk}=\sqrt{w}h_{mk}$ is an element of matrix $\mathbf{H}'$ and $w$ is a relaxation parameter. We let $\gamma={{{{\hat {\sigma} }^2}} \mathord{\left/
 {\vphantom {{{{\hat {\sigma} }^2}} {{{\tilde {\sigma} }^2}}}} \right.
 \kern-\nulldelimiterspace} {{{\tilde {\sigma} }^2}}}$ and assume $v_{k\to m}^v(t)$ and $v_{m\to k}^s(t)$ converges to $\hat{\sigma}^2$ and $\tilde{\sigma}^2$ respectively. We deal with the case when $\beta=K/M<1$.
 \subsection{SA-GMPID Algorithm}
 The message update of SA-GMPID of variable node (\ref{e3}) is
\begin{equation}\label{FC1}
\!\!\!\!\left\{\!\!\!\! {\begin{array}{*{20}{l}}
{v_{k \to m}^v(t) = {{\left( {\sum\limits_{i } {{h}_{ik}^2v_{i \to k}^{s{\;^{ - 1}}}(t) + \sigma _{{x_k}}^{ - 2}\;} } \right)}^{ - 1}},}\\
{e_{k \to m}^v\!(t) \!\!=\!\! v_{k \to m}^v(t)\!\!\sum\limits_{i} {{{h'}_{ik}}\!v_{i \to k}^{s{\;^{ - 1}}}\!(t)e_{i \to k}^s(t)\!-\!(\!w\!-\!1\!)e_{k \to m}^v\!(\!t\!-\!1\!).\;\;} }
\end{array}} \right.\quad
\end{equation}
The message update at the sum node (\ref{e2}) is changed as
\begin{equation}\label{FC2}
\left\{ \begin{array}{l}
e_{m \to k}^s(t) = {y'_m} - \sum\limits_{i \ne k} {{h'}_{mi}e_{i \to m}^v(t-1),} \\
v_{m \to k}^s(t)= \sum\limits_{i \ne k} {{h}_{mi}^2v_{i \to m}^v(t-1) + {{\sigma} _n^2}\;}.
\end{array} \right.
\end{equation}
After the iteration between (\ref{FC1}) and ({\ref{FC2}}), output
\begin{equation}\label{FC3}
\left\{ \!\!\!\begin{array}{l}
{\sigma}^2_{\hat{x}_k} = {\left( {\sum\limits_{m } {{h}_{mk}^2v_{m \to k}^{s{\;^{ - 1}}}(t) + \sigma _{{x_k}}^{ - 2}\;} } \right)^{ - 1}},\\
\hat{x}_k = \hat{\sigma}^2_{x_k}\sum\limits_{m } \left({{h'}_{mk}v_{m \to k}^{s{\;^{ - 1}}}(t)e_{m \to k}^s(t)\!-\!\frac{w-1}{M}e_{k \to m}^v(t-1)}\right)\!,
\end{array} \right.\quad
\end{equation}
where $k \!\in \!\{ 1,2, \cdots ,K\}$. Next, its convergence will be proved.

\subsection{Convergence Analysis of SA-GMPID}
%\begin{figure}
%  \centering
%  \includegraphics[width=8.0cm]{100_600Combine.pdf}\\
%  \caption{Performance comparison between the matched filter, MMSE detection and GMPID with 10 iterations; MMSE detection performance estimation and the proposed GMPID performance estimation. The simulations are for $100\times600$ MU-MIMO system with $\beta=1/6$.}\label{f4}
%\end{figure}

\textbf{\emph{Theorem 3:}}  \emph{When $\beta= K/M<1$ is fixed and $K\rightarrow \infty$, the SA-GMPID converges to the MMSE detector if the relaxation parameter $w$ satisfies $0<w<2/\lambda_{max}^\mathbf{A}$, where $\lambda_{max}^\mathbf{A}$ is the largest eigenvalue of $\mathbf{A}=\gamma\left(\mathbf{H}^T\mathbf{H}-\mathbf{D}_{\mathbf{H}^T\mathbf{H}}\right)+\mathbf{I}_{K}$.}

\begin{IEEEproof}
 The proof is omitted due to the page limit.
\end{IEEEproof}

The optimal relaxation parameter $w$ is given by
\begin{equation}\label{FC10}
w=2/({\lambda}_{min}^\mathbf{A}+{{\lambda}_{max}^\mathbf{A}}).
\end{equation}
By the random matrix theory \cite{verdu2004}, when $\beta= K/M$ is fixed and $K\rightarrow \infty$, the smallest and largest eigenvalues ${\lambda}_{max}^\mathbf{A}$ and ${\lambda}_{min}^\mathbf{A}$ of matrix $\mathbf{A}$ are estimated given by
\begin{equation}\label{eigv1}
{\lambda}_{min}^\mathbf{A}\!\!=\!1+\gamma M{\!\left[\!\left( \!{1 \!-\!\! \sqrt {\beta} } \right)^2\!\!\!-\!1\right]} ,{\lambda}_{max}^\mathbf{A}\!\!=\!1+\gamma M{\!\left[\!\!\left( \!{1\! +\!\! \sqrt {\beta} } \right)^2\!\!\!\!-\!1\right]}\!.
\end{equation}
The spectral radius of $\mathbf{I}_{K}-w\mathbf{A}$ is $\rho_{min}(\mathbf{I}_{K}-w\mathbf{A})= \frac{{\lambda}_{max}^\mathbf{A}-{\lambda}_{min}^\mathbf{A}}{{\lambda}_{max}^\mathbf{A}+{\lambda}_{min}^\mathbf{A}}<1$ . From (\ref{eigv1}), we have $w = {1 \mathord{\left/
 {\vphantom {1 {\left( {1 + \gamma M \beta} \right)}}} \right.
 \kern-\nulldelimiterspace} {\left( {1 + \gamma M \beta} \right)}}$. When $\beta<1$, from (\ref{p8}), $M\gamma\to 1$. Thus,
 \begin{equation}\label{FC12}
 w={1 \mathord{\left/
 {\vphantom {1 {(1 + \beta )}}} \right.
 \kern-\nulldelimiterspace} {(1 + \beta )}},\quad \rho_{min}(\mathbf{I}_{K}-w\mathbf{A})\approx\frac{2\sqrt{\beta}}{1+\beta}<1.
 \end{equation} Comparing with (\ref{radius1}), we get the following corollary.

\textbf{\emph{Corollary 2:}} \emph{The SA-GMPID converges faster than the GMPID when $\beta= K/M<1$ is fixed and $K\rightarrow \infty$.}
%\begin{figure}
%  \centering
%  \includegraphics[width=8cm]{Cov100_600.pdf}\\
%  \caption{Convergence illustration of GMPID. The curves from top and bottom are the performance for $100\times600$ MU-MIMO system with $1 \sim 10$ iterations.}\label{f6}
%\end{figure}

%However, for a finite $K$ and $M$, the smallest and largest eigenvalues of the matrix $A$ given by (\ref{eigv1}) are inaccurate, which may impact the converge rate and MSE performance of the FC-GMPID. Similarly, we can set the the relaxation parameter as $w=2/\lambda^*_A$ to improve the convergence speed and the robust of the algorithm and with a little more calculations as the cost, where $\lambda^*_A$ is an upper bound of eigenvalues \cite{Varga2009}
%\begin{equation}\label{FC14}
%{\lambda _{\max }} \le \lambda^*_A= \min \{ \mathop {max}\limits_{1 \le i \le K} \sum\limits_{j = 1}^K {\left| {{a_{ij}}} \right|} ,\mathop {max}\limits_{1 \le j \le K} \sum\limits_{i = 1}^K {\left| {{a_{ij}}} \right|} \}.
%\end{equation}

\section{Simulation Results}

\begin{figure}
  \centering
  \includegraphics[width=9cm]{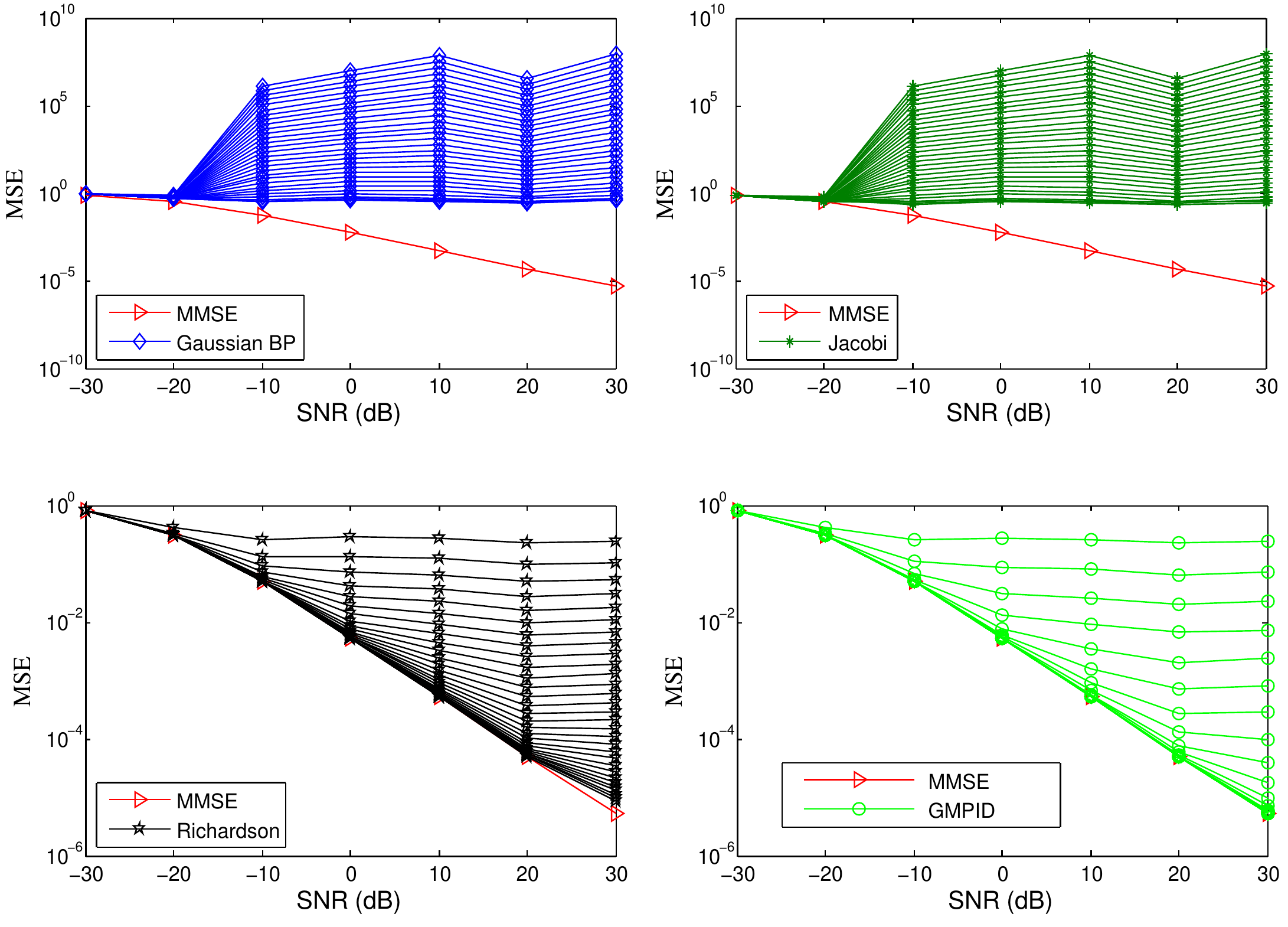}\\
  \caption{Performance comparison between GMPID and the other iterative methods: Gaussian BP, Jacobi and Richardson method. The simulations are for $100\times300$ MU-MIMO system with $\beta=1/3$ and $1\sim30$ iterations.}\label{f8}
  \vspace{-0.25cm}
\end{figure}
\begin{figure}
  \centering
  \includegraphics[width=9cm]{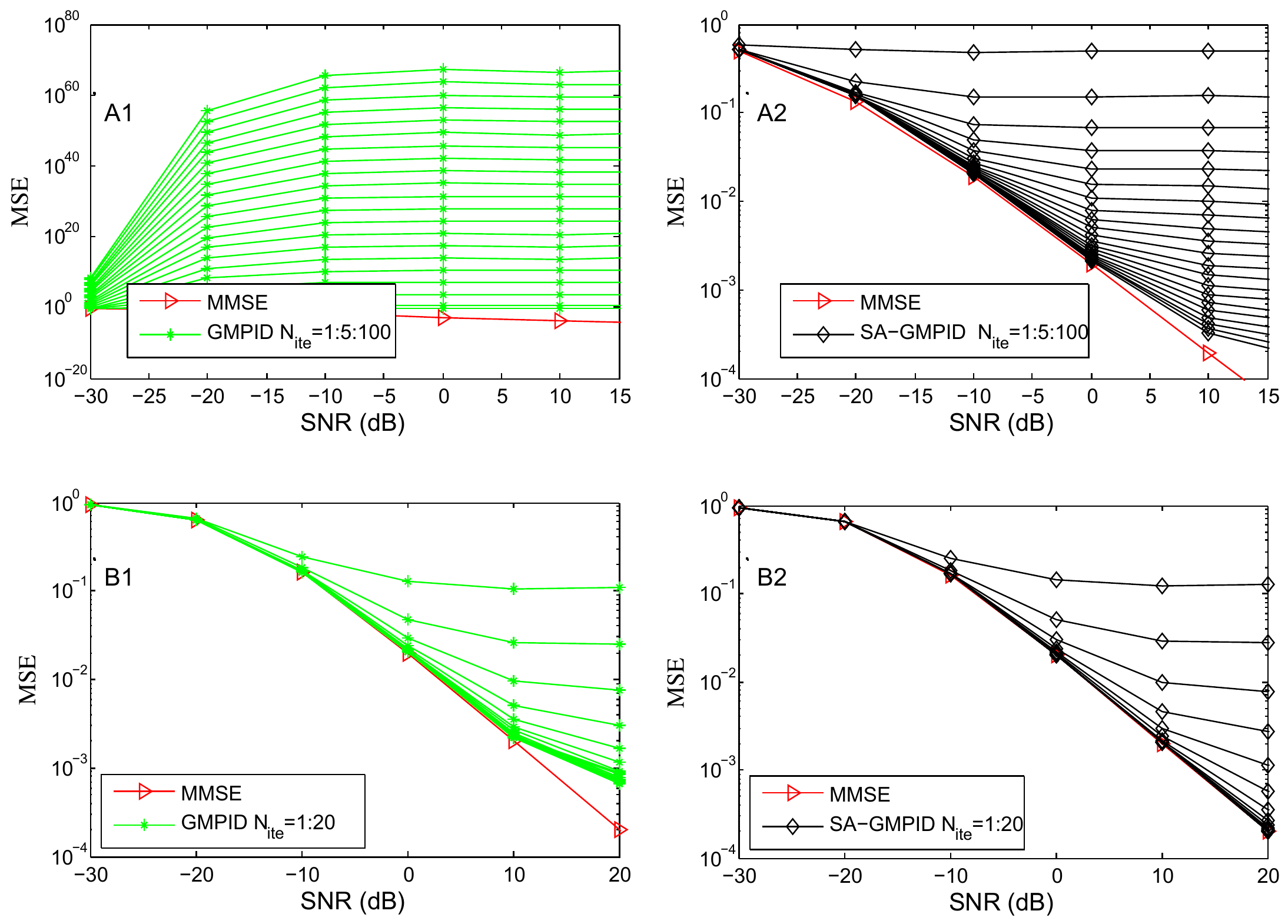}\\
  \caption{Performance comparison between the GMPID and SA-GMPID. Figures A1 and A2 are for $1000\times1500$ MU-MIMO system with $1\sim100$ iterations and $\beta=2/3$. Figures B1 and B2 are for $10\times60$ MU-MIMO system with $20$ iterations and $\beta=1/6$. }\label{f9}
  \vspace{-0.25cm}
\end{figure}
\newcommand{\tabincell}[2]{\begin{tabular}{@{}#1@{}}#2\end{tabular}}
\begin{table}[!t]
\renewcommand{\arraystretch}{1.4}
\caption{Convergence comparison between SA-GMPID, GMPID, Jacobi, GaBP and Richardson method. The ``C" and ``D" denote convergent and divergent respectively, and $``+"$ and $``-"$ denote the right limit and the left limit respectively.}
\label{table}
\centering
\begin{tabular}{|c|c|c|c|c|c|}
\hline
 Figure & $\beta$ & \tabincell{c}{ Jacobi\\ \& GaBP} &  GMPID  & \tabincell{c}{Richardson\\ \& SA-GMPID} \\
\hline
  Fig. 4 &$\beta<(\sqrt{2}-1)^2$ & C & C & C \\
\hline
 Fig. 5 & $\beta\to(\sqrt{2}-1)^2_+$ & D & C & C \\
\hline
 Fig. 6 & $\beta\to1_-$ & D & D & C \\
\hline
\end{tabular}
\vspace{-0.3cm}
\end{table}
In this section, we give the numerical results of the proposed GMPID for the MU-MIMO system with Gaussian sources. Assume that the sources are i.i.d. with $x_k\sim\mathcal{N}(0,1)$ and the entries of the channel matrix $\mathbf{H}$ are i.i.d. with normal distribution $\mathcal{N}(0,1)$. In the following simulations, $N_{ite}$ denotes the number of iterations, $SNR=\frac{1}{\sigma^2_n}$ and $MSE=\frac{1}{K}(\textbf{\emph{x}}-\hat{\textbf{\emph{x}}})^2$ denotes the average mean squared error. All the simulations are repeated 500 times to get the results.

%Fig. \ref{f4} presents the average MSE performance and the performance estimations of the different detection methods for the MU-MIMO system with $K=100$, $M=600$ and $\beta=1/6$. The curve \emph{RT Estimation} is the MMSE detection performance estimation given by the random matrix theory and the curve \emph{GMPID Estimation} is the GMPID performance estimation given by (\ref{p6}). It can be seen that the performance of MF (Matched Filter) is terrible, and the other curves are almost completely overlapped. These verify that 1) the proposed GMPID converges fast (only 10 iterations) to the MMSE detection (Theorem 2),  2) the performance estimations with random matrix theory and GMPID match well to the simulation results (Theorem 1) and 3) the complexity of GMPID is much less than that of the MMSE detection ($\mathcal{O}(K^2M+K^3)$).

Fig. \ref{f8} gives the average MSE performance and convergence comparison between the SA-GMPID and the other iterative methods: Jacobi method, Gaussian BP method and Richardson method \cite{Gao2014}, where $K=100$ and $M=300$. We can see that the GMPID converges faster (although $\beta=1/3>(\sqrt{2}-1)^2$ ) to the MMSE detection than the other three methods. Furthermore, the SA-GMPID converges even when the Jacobi method and Gaussian BP method diverge. It should be noted that SA-GMPID has the lowest complexity.

Fig. \ref{f9} gives the average MSE performance comparison between the GMPID and SA-GMPID for the cases that $\beta=2/3$, $K=1000$, $M=1500$ with $100$ iterations (figures A1 and A2) and $\beta=1/6$, $K=10$, $M=60$ with 20 iterations (figures B1 and B2), respectively. We can see that the GMPID is diverging when $\beta\to1$ with figure A1. In particular, figure A2 shows the SA-GMPID converges to the MMSE detection with the increase number of iterations. This verifies our analysis result in Theorem 3. Furthermore, figures B1 and B2 show that 1) SA-GMPID converges faster to the MMSE detection than GMPID (Corollary 2) and 2) the proposed theoretical results are also suitable for MU-MIMO systems with a small number of antennas and users.

Table \ref{table} concludes the convergence comparison of the different detection methods, where ``C" (or ``D") denotes convergent (divergent) and $``+"$ (or $``-"$) denotes right limit (left limit). It shows that 1) all the iterative methods are convergent when $\beta<(\sqrt{2}-1)^2$, 2) the Jacobi and GaBP methods are divergent when $\beta$ larger than $(\sqrt{2}-1)^2$, 3) GMPID is still convergent when $\beta$ close to $(\sqrt{2}-1)^2_+$ and 4) Richardson method and SA-GMPID are convergent when $\beta$ close to $1$.
\section{Conclusion}
A low-complexity detection method GMPID has been discussed, in which the means and variances are transmitted between the variable nodes and sum nodes. The convergence of GMPID has been analysed. It is proved that the variances of GMPID converge to the MSE of MMSE detection. Two sufficient conditions that the GMPID converges to the MMSE detection have also been presented. As the GMPID does not converge when $\beta\to1$, SA-GMPID has been proposed, which converges to MMSE detection in any case that $K<M$. Numerical results are provided to verify the proposed theoretical results. It should be noted that our simulations show that the proposed theories are also suitable for MU-MIMO systems with a limited number of antennas and users.

\end{document}